\documentclass[11pt]{article}
\usepackage{moriondmike,epsfig,graphics,color,axodraw}

\def\be{\begin{equation}}
\def\ee{\end{equation}}
\def\bea{\begin{eqnarray}}
\def\eea{\end{eqnarray}}

\begin{document}
\hfill\begin{minipage}{5cm}
hep-ph/0007051\\
July 2000
\end{minipage}
\vspace*{4cm}
\def\thefootnote{\fnsymbol{footnote}}
\begin{center}JETS IN HADRON COLLISIONS~\footnote{Talks given at XXXVth
Rencontres de Moriond, QCD and Hadronic Interactions, Les Arcs 1800,
France, March 18th--25th 2000, and DIS 2000, 8th International Workshop
on Deep-Inelastic Scattering, Liverpool, UK, April 25th--30th 2000.}
\end{center}
\def\thefootnote{\alph{footnote}}

\vspace*{1cm}
\centerline{Michael H. Seymour}

\vspace*{1cm}
\begin{center}Theoretical Physics Group, Particle Physics Department,\\
  CLRC~Rutherford Appleton Laboratory,\\
  Chilton, Didcot.  OX11~0QX.  United~Kingdom.
\end{center}

\vspace*{1cm}
\centerline{\bf Abstract}
\begin{quote}
We discuss recent progress and open questions in QCD jet physics, with
particular emphasis on two areas: jet definitions and jet substructure.
\end{quote}
\newpage

\vspace*{4cm}
\title{JETS IN HADRON COLLISIONS}

\author{M.H. SEYMOUR}

\address{Theoretical Physics Group, Particle Physics Department,\\
  CLRC~Rutherford Appleton Laboratory,\\
  Chilton, Didcot.  OX11~0QX.  United~Kingdom.}

\maketitle\abstracts{
We discuss recent progress and open questions in QCD jet physics, with
particular emphasis on two areas: jet definitions and jet substructure.
}

\section{Introduction}

Jet physics is a very vibrant field and there has been a great deal of
progress both experimental and theoretical in the last few years.
Rather than trying to give a review of the whole field, I have chosen to
go into two topics in more depth: jet definitions and jet substructure.

\section{Jet definitions}

\subsection{Cone algorithms}

The standard jet algorithms used by most hadron-collider experiments
have been based on the geometric cone definition.  Although the general
idea of this definition is straightforward, when implementing it one is
faced with myriad choices and historically each experiment implemented
its own algorithm.  The Snowmass Accord was an attempt to unify these
and agree on one definition that theorists and experiments could use.
It defines jets by finding directions in rapidity--azimuth,
$\eta\!-\!\phi$, space that maximize the amount of hadronic energy
flowing through a cone of fixed radius (in $\eta\!-\!\phi$ space), $R$,
drawn around them.  The jet momentum is defined to be massless with
transverse energy, $E_T$, rapidity and azimuth calculated from those of
the particles in the cone, as
\begin{eqnarray}
  \label{snow1}
  E_T &=& \sum_{i\in\mathrm{cone}} E_{Ti}, \\
  \eta &=& \frac1{E_T}\sum_{i\in\mathrm{cone}} E_{Ti}\eta_i, \\
  \phi &=& \frac1{E_T}\sum_{i\in\mathrm{cone}} E_{Ti}\phi_i.
  \label{snow3}
\end{eqnarray}

\subsubsection{Iterative cone algorithms}

The CDF experiment, which first tried to implement the Snowmass
Accord\cite{CDF1}, found that it was not a complete definition and had
to be supplemented for two reasons.  The result was an iterative cone
algorithm.  Subsequent experiments have followed a similar line,
although the fine details vary from experiment to experiment.

The first problem is that the maximization process was not uniquely
defined.  A global maximization proved too costly in computer time to be
practical, so they defined a local maximization which was achieved
iteratively.  They define a set of directions to be \emph{seed
directions}, draw a cone around each and apply
Eqs.~(\ref{snow1}--\ref{snow3}), which define a new direction.  A cone
is drawn around this direction and Eqs.~(\ref{snow1}--\ref{snow3}) used
again iteratively, until a stable direction is obtained.  It can be
shown that provided all calorimeter cells have a positive energy (which
is not necessarily the case experimentally, for example in D0) a stable
direction is always reached and that it gives a local maximum of the
energy in the cone.  The definition of the seed directions is closely
tied to the details of specific detectors, but is typically every
calorimeter cell above some energy threshold, eg 1~GeV.

The second problem is that the jets so defined often overlap and share
energy in common, while a mapping of each calorimeter cell to only one
jet was sought, so a merging/splitting algorithm was added.  Again the
precise details vary from experiment to experiment, but the general idea
is to either merge the two overlapping jets into one, if the overlap
region contains more than a given fraction of their total energy, or to
split them into two along the half-way line, otherwise.

It has recently been realized\cite{GK,S1} that the iterative cone
algorithm is not infrared safe.  The problem is that the iteration from
the seed directions is not exhaustive: it is not guaranteed to produce a
complete list of all local maxima.  If two cones overlap in such a way
that their centres can also be enclosed in one cone but there is little
energy in the overlap region, then it turns out that the outcome is
different depending on whether or not the overlap region contains a seed
direction.  This results in a logarithmic dependence on the seed cell
threshold which would give a divergent cross section, if the threshold
were taken to zero for the purposes of making an idealized calculation.
This divergence first shows up when there can be three nearby partons,
which for jets in hadron collisions is NLO in the three-jet cross
section\cite{GK} and NNLO in the two-jet or inclusive one-jet cross
sections\cite{S1}.

It is also worth noting that this is mainly a problem for order-by-order
perturbation theory.
\begin{figure}[t]
\centerline{\resizebox{!}{0.33\textwidth}{\includegraphics[76,230][523,566]{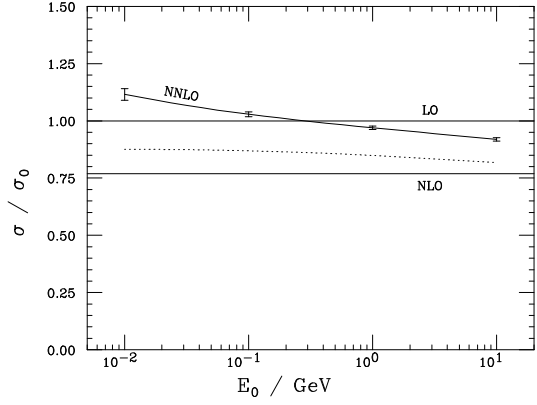}}}
\caption{The seed cell threshold dependence of the inclusive jet cross
section in the D0 jet algorithm with $R=0.7$ in fixed-order (solid) and
all-order (dotted) calculations.  Taken from~\protect\cite{S1}.}
\label{Eseed1}
\end{figure}
As shown in Fig.~\ref{Eseed1} (Fig.~2 of~\cite{S1}), after summing to
all orders, the dependence on the cutoff is very weak.  This is because
physically almost every such event does in fact contain a seed direction
and one gets a Sudakov form factor much less than unity.  When expanded
order-by-order in $\alpha_s$, such a form factor gives large terms at
every order.

The solution\cite{E,S1} is to add an additional stage of iteration.
After the first stage, but before the merging/splitting algorithm,
additional seed directions are tried, defined as the mid-points of any
overlapping cones.
\begin{figure}[t]
\centerline{\resizebox{!}{0.33\textwidth}{\includegraphics[76,230][523,566]{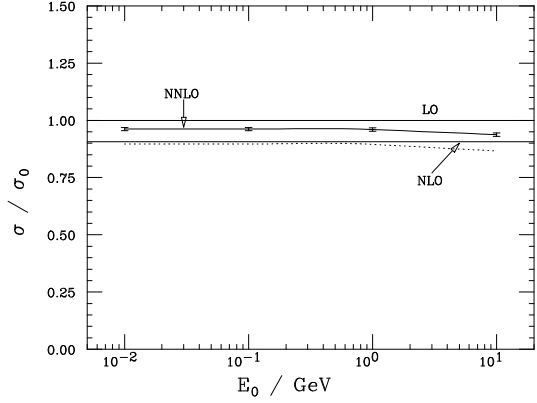}}}
\caption{The seed cell threshold dependence of the inclusive jet cross
section in the improved iterative cone algorithm, in which mid-points of
pairs of overlapping jets are used as additional seeds for the
jet-finding, with $R=0.7$ in fixed-order (solid) and
all-order (dotted) calculations.  Taken from~\protect\cite{S1}.}
\label{Eseed2}
\end{figure}
As shown in Fig.~\ref{Eseed2} (Fig.~4b of~\cite{S1}), this results in
much more stable cross sections, which are finite order-by-order in
perturbation theory.

\begin{figure}[t]
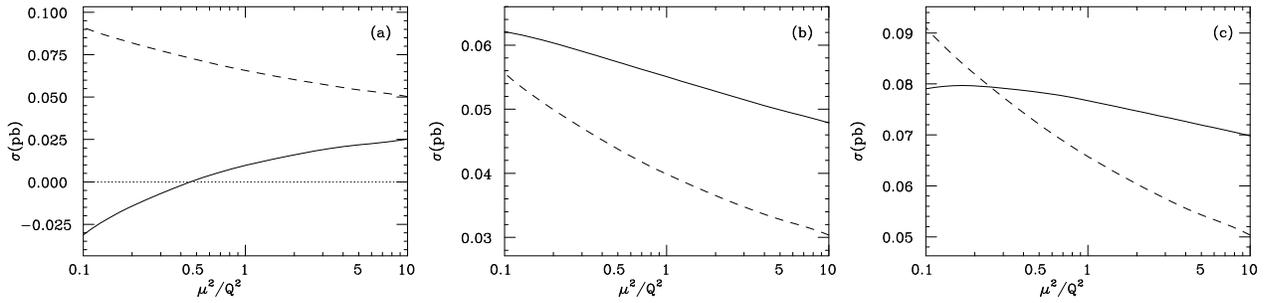

\centerline{%
\hspace*{-0.02\textwidth}%
\resizebox{0.34\textwidth}{!}{\includegraphics{bjorn_01.ps}}%
\hfill%
\resizebox{0.34\textwidth}{!}{\includegraphics{bjorn_02.ps}}%
\hfill%
\resizebox{0.34\textwidth}{!}{\includegraphics{bjorn_03.ps}}%
\hspace*{-0.02\textwidth}%
}
\caption{The two-jet cross section at high $Q^2$ in the HERA lab frame
  at LO (dashed) and NLO (solid) in the CDF cone algorithm (a), the
  improved cone algorithm (b) and the $k_\perp$ algorithm (c).  Taken
  from~\protect\cite{S2}.}
\label{bjorn}
\end{figure}

Although it was originally thought that, as stated above, this only
became important to the inclusive cross section at NNLO, it has more
recently been realized\cite{S2,R2} that in DIS it appears at NLO, if the
jets are analyzed in the lab frame.  This is because the outgoing
electron acts kinematically like a jet, against which the other jets in
the event recoil, but since it is not coloured it does not contribute to
the QCD corrections.  We can therefore test these ideas using standard
NLO calculations of two-jet production in DIS.  An example is shown in
Fig.~\ref{bjorn} (Fig.~1 of~\cite{S2}), for the dijet cross section in
the lab frame at HERA.
The results in the iterative cone algorithm are clearly out of control,
while those in the improved cone algorithm are considerably better.  It
is worth noting that the $k_\perp$ algorithm, to be discussed shortly,
is better behaved still.  Similar results were found by the jets working
group of the Physics at Run~II workshop\cite{R2}.

\subsubsection{The Improved Legacy Cone Algorithm}

A recent innovation arising from the Physics at Run~II workshop is a new
accord on how to define cone jets, the ILCA, based on the improvement
suggested in Ref.~\cite{E,S1}.
\begin{figure}[t]
\centerline{\hfill%
\scalebox{0.5}{\includegraphics[149,109][397,607]{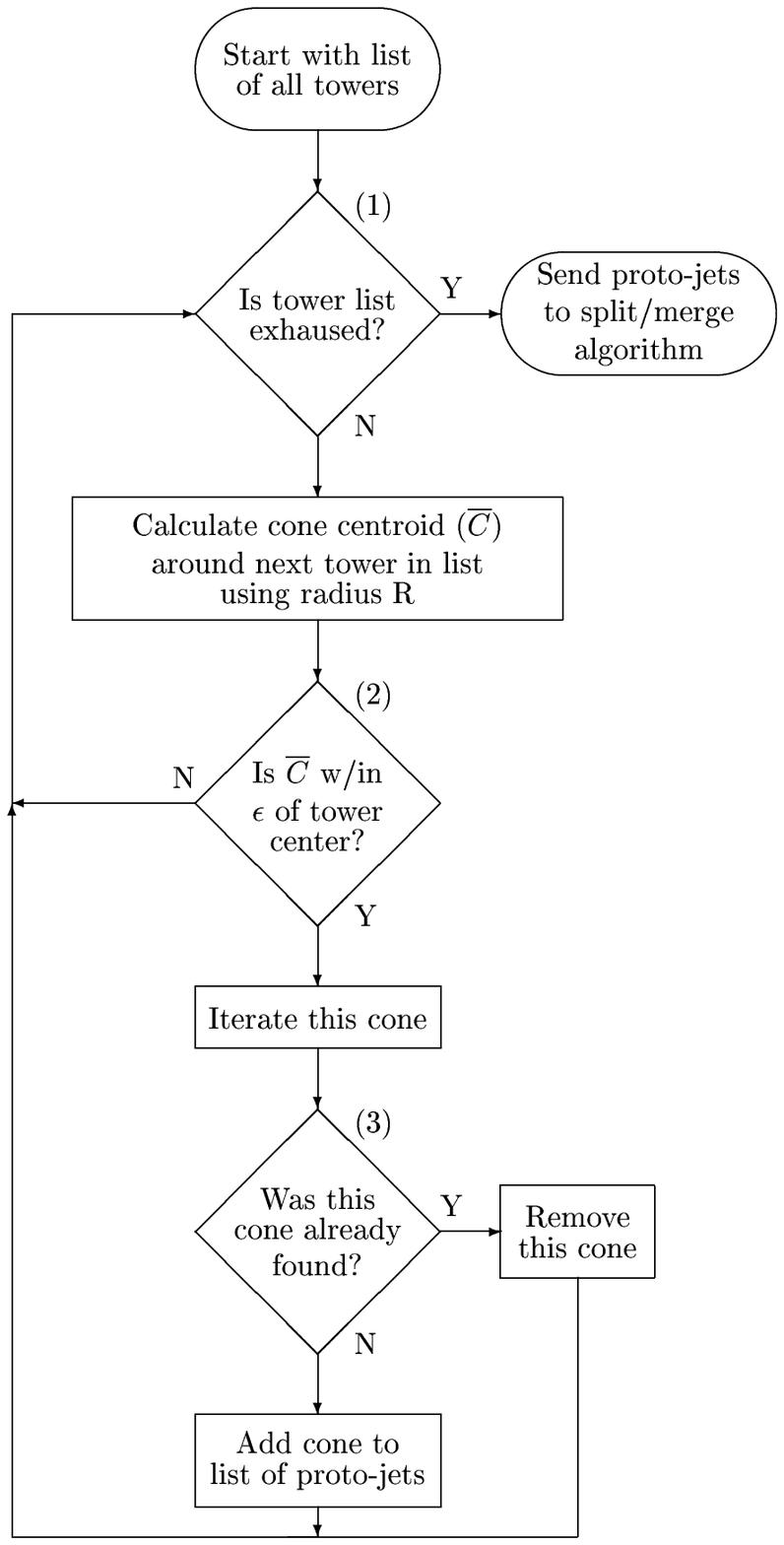}}\hfill%
\scalebox{0.5}{\includegraphics[234,338][375,666]{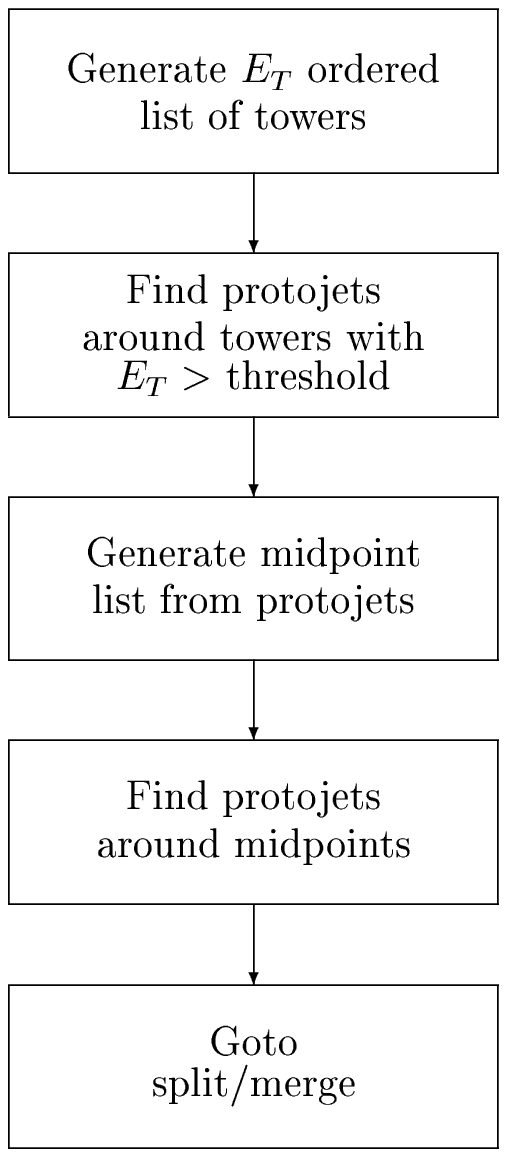}}\hfill%
\scalebox{0.5}{\includegraphics[147,122][463,666]{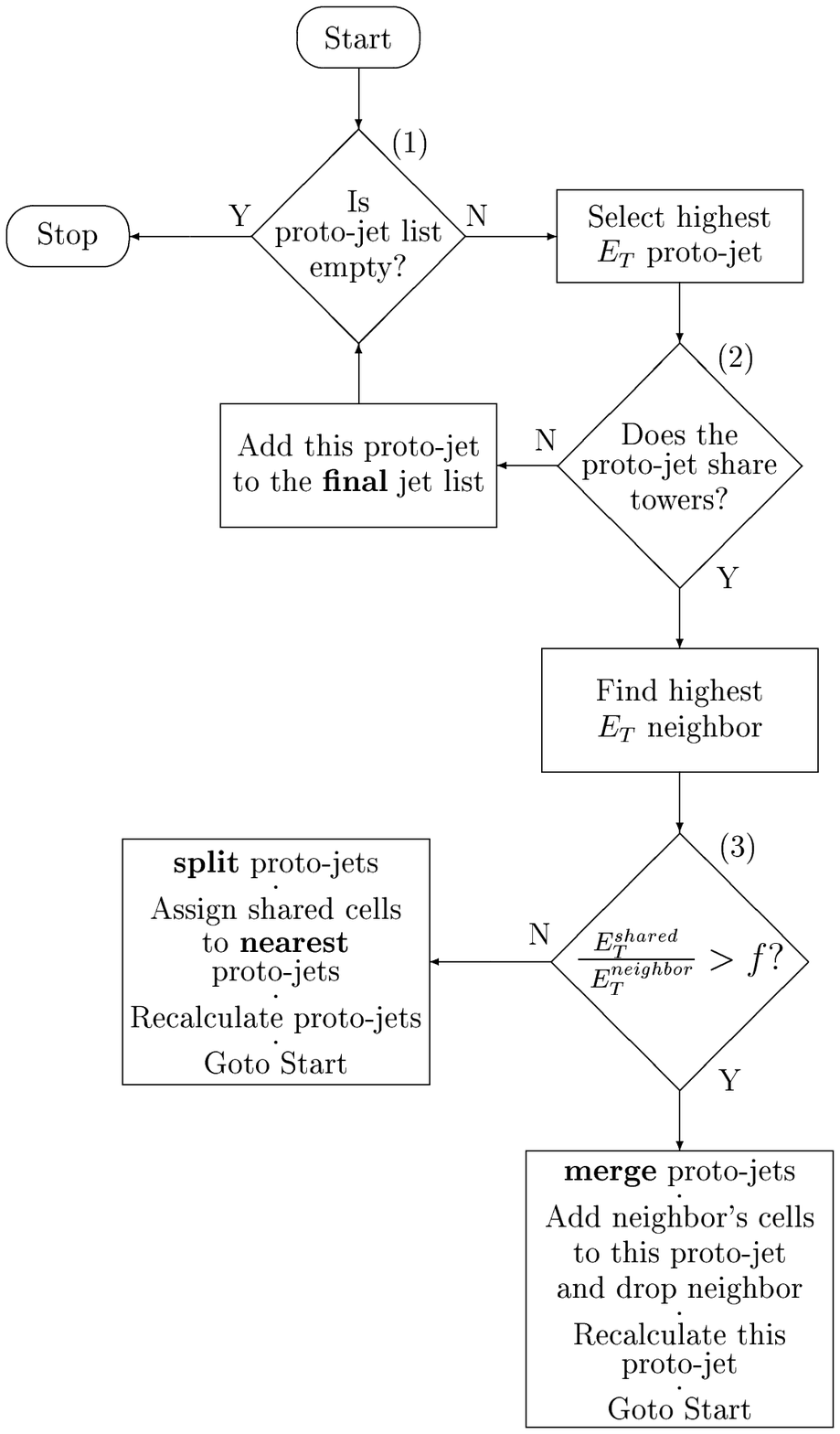}}%
\hfill}
\caption{The full specification of the Improved Legacy Cone Algorithm.
Taken from~\protect\cite{R2}.}
\label{ILCAflo}
\end{figure}
As shown in Fig.~\ref{ILCAflo} (Figs.~4, 5 and~18 of~\cite{R2}) it is
fully algorithmic, if a little cumbersome, meaning that any experiment
or theorist can implement it in exactly the same way.  Among the
requirements it had to fulfil is that its numerical result be within
5\% of the algorithms used by CDF and D0 in Run~I, which is the case.
Despite this small difference at the hadron level, it is finite
order-by-order in perturbation theory and has smaller hadronization
corrections.  It is therefore a significant step forwards.

\subsection{The $k_\perp$ cluster algorithm}

Despite the improvements in the cone algorithm, it still has problems
relative to the $k_\perp$ cluster algorithm\cite{CDSW,ES}.
\begin{figure}[t]
\centerline{\scalebox{0.5}{\includegraphics[105,33][469,666]{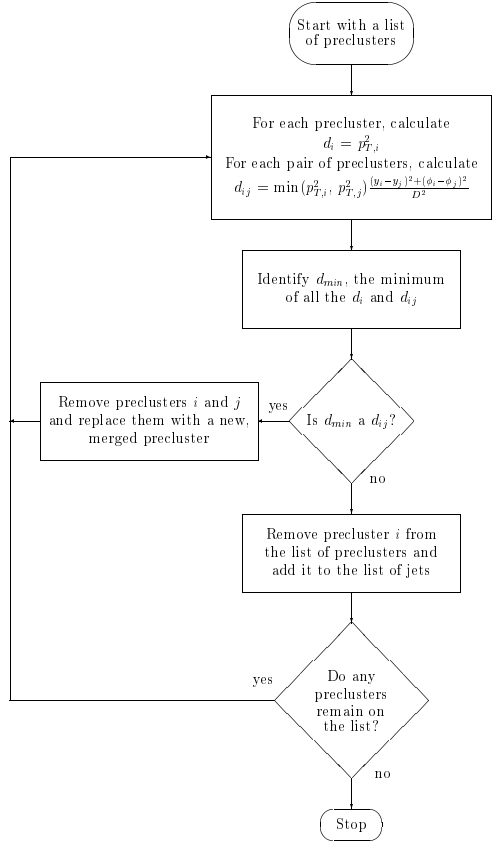}}}
\caption{The full specification of the inclusive $k_\perp$ algorithm.
Taken from~\protect\cite{R2}.}
\label{ktflo}
\end{figure}
The definition of this is shown in Fig.~\ref{ktflo} (Fig.~19
of~\cite{R2}) in the same notation: it is clearly much simpler.  Its
results depend on an input parameter $R$ (sometimes called~$D$), which
actually plays a similar role to the radius parameter in the cone
algorithm.  Among its advantages are its simplicity, the fact that it
exhaustively maps every hadron in the final state to one and only one
jet with no overlaps, and the fact that it is based on the $k_\perp$
measure, allowing the phase space for sequential soft gluon emission
to be factorized and the corresponding large logarithms to be summed
to all orders.  It also suffers smaller hadronization and detector
corrections than the cone algorithm in practice.

\begin{figure}[t]
\centerline{\resizebox{!}{0.33\textwidth}{\includegraphics[21,10][437,280]{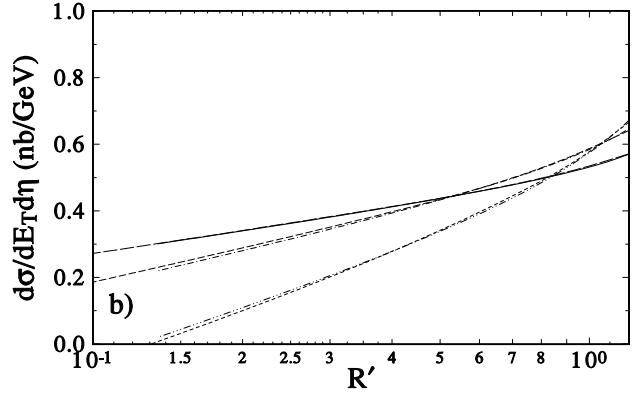}}}
\caption{Order $\alpha_s^3$ inclusive jet cross section for
$E_T=100$~GeV, $\sqrt{s}=1800$~GeV averaged over $\eta_J$ in the range
$0.1<|\eta_J|<0.7$.  The pairs of curves corresponding to the two
algorithms are: $\mu=E_T$ (solid), $\mu=E_T/2$ (dot-dash), $\mu=E_T/4$
(dot-dot-dot-dash) plotted against $R'=1.35R_{cone}$ for the cone
algorithm and against $R'=R_{comb}$ for the $k_\perp$ algorithm.  Taken
from~\protect\cite{ES}.}
\label{ES1}
\end{figure}

\begin{figure}[t]
\centerline{\resizebox{!}{0.33\textwidth}{\includegraphics[21,10][437,280]{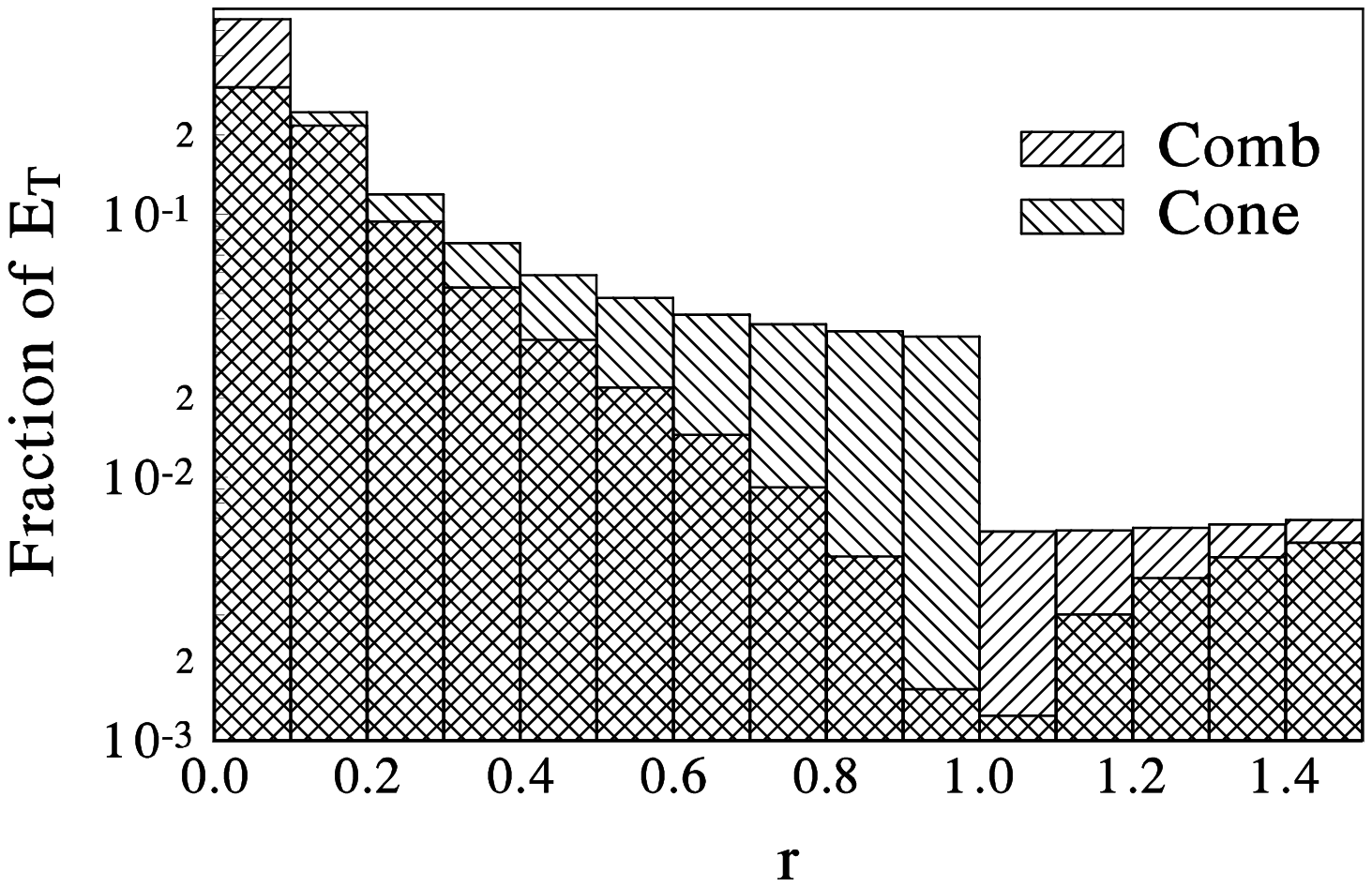}}}
\caption{The fraction of jet $E_T$ in angular annuli $r$ to $r+0.1$
comparing the cone and $k_\perp$ (``comb'') jet algorithms.  In both
cases the jet has $R=1.0$, $E_T=100$~GeV, $\sqrt{s}=1800$~GeV,
$0.1<|\eta_J|<0.7$ and renormalization/factorization scale $\mu=E_T/2$.
Taken from~\protect\cite{ES}.}
\label{ES2}
\end{figure}

\begin{figure}[t]
\centerline{\rotatebox{90}{\resizebox{0.33\textwidth}{!}{\includegraphics[100,104][522,660]{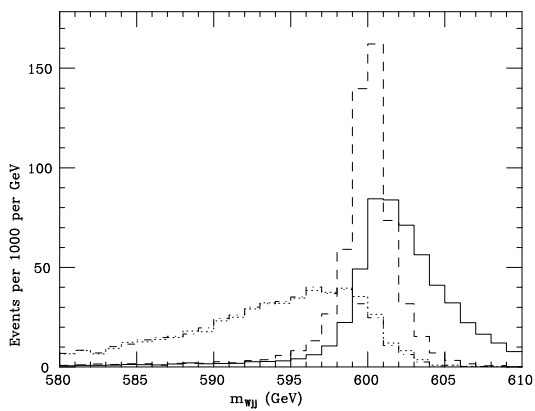}}}}
\caption{Reconstructed mass distribution of Higgs candidates (with a
fixed Higgs mass of 600~GeV) according to the cluster (solid and dashed)
and cone (dotted and dot-dashed) algorithms, at calorimeter level with
(solid and dotted) and without (dashed and dot-dashed) particles from
the underlying event.  Taken from~\protect\cite{S4}.}
\label{Higgs}
\end{figure}

As shown in Fig.~\ref{ES1} (Fig.~2b of~\cite{ES}), at the level of the
NLO inclusive jet cross section, the two jet definitions are essentially
identical.
However, even at that order the energy spread within the jet is quite
different, as shown in Fig.~\ref{ES2} (Fig.~1 of~\cite{ES}).
The cluster algorithm pays more attention to the core of the jet, while
the local optimization inherent in the cone algorithm does its best to
suck as much soft junk into the edges of the jet as possible.  This is
thought to be why the cluster algorithm gives significantly cleaner
reconstruction of highly boosted objects as shown in Fig.~\ref{Higgs}
(Fig.~1b of~\cite{S3}).

Both H1 and ZEUS now use the $k_\perp$ algorithm as their `algorithm of
choice', and CDF and D0 are planning to use it on an equal footing with
the ILCA in Run~II.

\subsubsection{Preclustering}

One small problem with the $k_\perp$ algorithm that has yet to be solved
is the possible need for a preclustering step.  This is needed by the
Tevatron experiments for several reasons.

Firstly in D0 some calorimeter cells have negative energy, and it is not
entirely clear whether these can be incorporated into the algorithm
(although it is not clear to me that they cannot, for example by
replacing $E_{Ti}^2$ by
\begin{equation}
  E^{\prime2}_{Ti} \equiv E_{Ti}^2 \times \mathrm{sign}(E_{ti}).
\end{equation}
These negative energy cells would then always be clustered with their
nearest neighbours in a process that would always continue until there
are no negative energy clusters left, regardless of the jet resolution
criterion).

Secondly, due to calorimeter segmentation and the finite transverse size
of hadronic showers, it is possible for one hadron to produce two or
more non-zero energy cells, or for two hadrons to shower into a single
cell.  Preclustering reduces the size of the detector corrections
associated with these effects, particularly at small subjet resolution
scales.

Finally, the clustering process takes ${\cal{O}}(n^3)$ time, where $n$
is the number of initial momenta.  For Tevatron events this time can be
prohibitive if all calorimeter cells are used as input.  It can be
reduced considerably by a little preclustering\cite{R2}.

Unfortunately no theoretical implementation of preclustering has been
proposed as yet.  It is not even clear whether it is possible to satisfy
the experimental needs with a theoretically-calculable algorithm.  A
possible solution is to run the inclusive $k_\perp$ algorithm with a
small $R$ parameter, $R\sim0.1\!-0.2$, and to use the output of this
algorithm as an input to the main algorithm.  It seems likely that the
large logarithms associated with this could be summed to all orders, but
it has not been explicitly checked.  Clearly it does not solve the
problem of cpu time, so is not a complete solution, but if it could be
shown that this has similar results to the experimental algorithm with
the same precluster radius then it could be used as a common ground on
which to compare theory and experiment.  That is, one could correct the
experimentally-preclustered calorimeter-level results to
theoretically-preclustered hadron level, rather than all the way to
un-preclustered hadron level.

\section{Jet structure}

\subsection{Jet shape}

The classic way to study the internal structure of jets is with the jet
shape.  This is inspired by the cone algorithm, although its use is not
limited to cone jets.  The jet shape\footnote{Note that, perversely, the
HERA experiments have defined their notation for $\Psi$ and $\rho$ to be
interchanged relative to the original definitions used by the Tevatron
experiments.  Here we use the HERA definition.} $\Psi(r)$ is defined as
the fraction of the jet's energy contained in a cone of radius $r$
centred on its direction.  We therefore have $\Psi(R)\sim1$, meaning
that the jet's energy is all contained within a cone of radius $R$ (the
relation is not exact because in neither the cluster algorithm, nor even
the cone algorithm after the merging/splitting step, is the edge of a
jet an exact geometric cone).  Narrower jets are characterized by larger
values of $\Psi(r)$.  The jet shape is sometimes discussed in terms of
the energy fractions in concentric angular annuli,
$\rho(r)\equiv-\mathrm{d}\Psi/\mathrm{d}r$.

\begin{figure}[t]
\centerline{\resizebox{!}{0.33\textwidth}{\includegraphics[14,26][509,283]{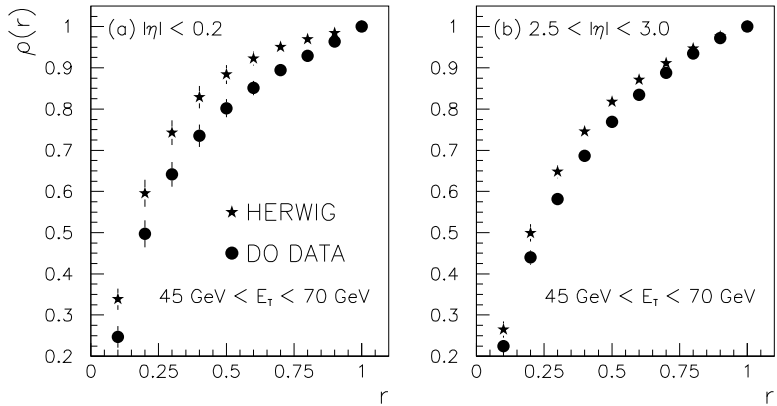}}}
\caption{The average integrated $E_T$ fraction versus the subcone radius
is plotted for the data and HERWIG Monte Carlo program, at calorimeter
level, for the $E_T$ range 45--70~GeV.  Taken from~\protect\cite{D01}.}
\label{D0_js}
\end{figure}

It has long been known that parton shower Monte Carlo programs like
HERWIG predict considerably narrower jets than are observed in the
Tevatron data, for example Fig.~\ref{D0_js} (Fig.~1 of~\cite{D01}).
Since jet shapes in $\mathrm{e^+e^-}$ annihilation were known to be well
described, even when separated out into quark- and gluon-jet
samples\cite{OPAL}, possible explanations focused on the two main new
ingredients in hadron collisions, initial-state radiation (ISR) and the
underlying event.  The ISR model is well tested by the Tevatron
experiments' measurements of colour coherence effects in two-jet events,
which HERWIG describes well\cite{CDF2}.  This leaves underlying event
effects as the most likely culprit.

\begin{figure}[t]
\centerline{\resizebox{!}{0.33\textwidth}{\includegraphics[5,5][555,365]{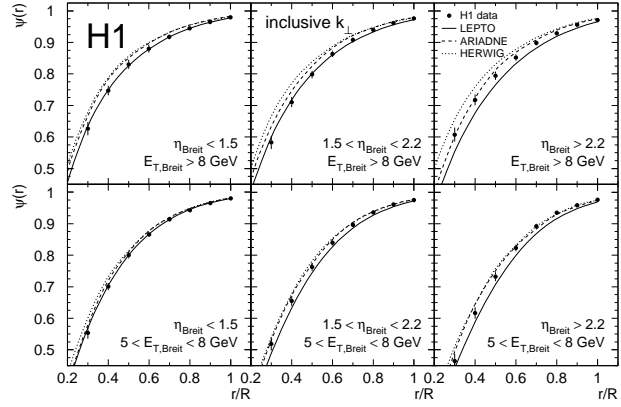}}}
\caption{The jet shapes for the inclusive $k_\perp$ algorithm.  The data
are shown as a function of the transverse jet energy and the jet
pseudo-rapidity in the Breit frame.  The results are compared to
predictions of QCD models.  Taken from~\protect\cite{H1}.}
\label{H1}
\end{figure}

\begin{figure}[t]
\centerline{\resizebox{!}{0.33\textwidth}{\includegraphics[5,5][257,231]{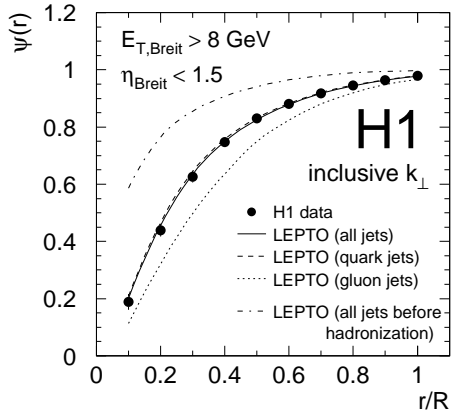}}}
\caption{Model predictions of the jet shape for the inclusive $k_\perp$
algorithm from the LEPTO parton shower model.  The jet shapes are shown
separately for quark and gluon induced jets with
$E_{T,\mathrm{Breit}}>8$~GeV and $\eta_{\mathrm{Breit}}<1.5$ together
with the sum of both and the comparison to the H1 measurement.  The
distribution before hadronization is also shown.  Taken
from~\protect\cite{H1}.}
\label{H2}
\end{figure}

This hypothesis can be tested at HERA, since resolved photoproduction
should have an underlying event and direct photoproduction and DIS
should not.  A great deal of excellent data have appeared in the last
couple of years\cite{H1,ZEUS,rest}, from which we choose just one
example, dijet events in DIS from H1, Fig.~\ref{H1} (Fig.~5
of~\cite{H1}).
It can be seen that HERWIG's prediction is again too narrow, although by
much less than at the Tevatron.  Since there should be no underlying
event correction, this clearly needs to be understood in more detail.
It is worth noting that in this kinematic region the hadronization
corrections are huge, as shown in Fig.~\ref{H2} (Fig.~6a of~\cite{H1})
for the LEPTO generator.

As mentioned earlier, DIS in the HERA lab frame has a special role in
jet physics because the recoiling electron acts kinematically like a
parton but is not coloured.  This has enabled the first NLO calculation
of the jet shape to be made\cite{KZ}.
\begin{figure}[t]
\centerline{\resizebox{!}{0.33\textwidth}{\includegraphics[39,48][468,522]{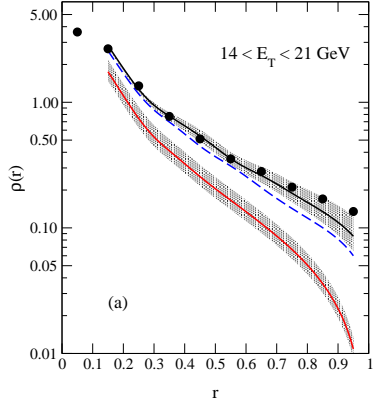}}}
\caption{Comparison of ZEUS jet shape data~\protect\cite{ZEUS} with QCD
predictions for DIS jets reconstructed by the iterative cone algorithm.
Jet cuts are: $-1<\eta<2$ and 14~GeV${}<E_T<{}$21~GeV.  ZEUS data
(circles) are compared with {\protect\textcolor{red}{LO (lower band)}}
and {\protect\textcolor{blue}{NLO (dashed line)}} QCD predictions.  The
upper band represents the NLO jet shape with additional cuts that are
not made on the data.  The width of the bands corresponds to varying the
renormalization scale between $\mu_r^2=\alpha_sQ^2/4$ and
$\mu_r^2=4\alpha_sQ^2$.  Taken from~\protect\cite{KZ}.}
\label{KZ}
\end{figure}
A comparison with ZEUS data is shown in Fig.~\ref{KZ} (Fig.~4a
of~\cite{KZ}).  For precisely the reasons mentioned earlier this is
extremely sensitive to the details of the jet algorithm.  Since the
iterative algorithm used by the ZEUS experiment is not infrared safe, it
cannot be used in the NLO calculation, so this comparison can only be
taken as indicative.  To supposedly take account of this, the authors
of~\cite{KZ} applied an additional cut in the calculation that was not
applied to the data and chose a very small scale for the running
coupling.  Bearing this in mind, and the fact that the hadronization
corrections shown in Fig.~\ref{H2} (Fig.~6a of~\cite{H1}) are about a
factor of two, the claimed good agreement shown in Fig.~\ref{KZ}
(Fig.~4a of~\cite{KZ}) must be seen as coincidental.

\subsection{Subjet studies}

\begin{figure}[t]
\centerline{\resizebox{!}{0.33\textwidth}{\includegraphics{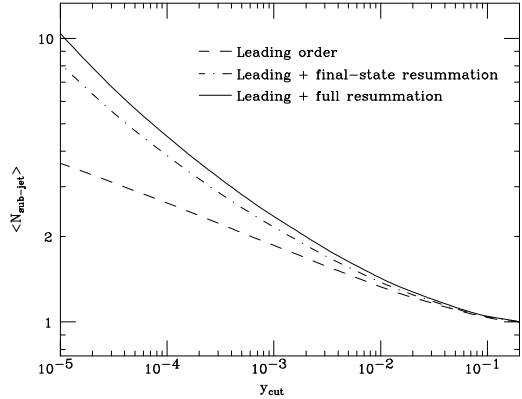}}}
\caption{The multiplicity of subjets in a 100~GeV jet according to the
  leading-order matrix element (dashed), matched leading-order and
  final-state logs (dot-dashed) and the full result with matched
  leading-order and leading and next-to-leading logs (solid).  Taken
  from~\protect\cite{S4}.}
\label{Ssub}
\end{figure}

The $k_\perp$ algorithm naturally suggests a new way of analysing the
internal structure of jets that is much closer to the partonic picture
of how that structure arises.  After identifying a jet of a given $E_T$,
we rerun the $k_\perp$ algorithm, but only on those particles that were
assigned to this jet.  We stop clustering when all values of $d_{ij}$
satisfy $d_{ij}>y_{\mathrm{cut}}E_T^2$, namely when all internal
relative transverse momenta are greater than
$\sqrt{y_{\mathrm{cut}}}E_T$.  Analysing the jets in this way is
extremely similar to analysing $\mathrm{e^+e^-}$ annihilation events at
$\sqrt{s}=E_T$ and the same value of $y_{\mathrm{cut}}$ and in fact it
can be shown\cite{S4} that the leading logarithms are identical.  Even
the ISR of gluons into the jet, which contributes at next-to-leading
logarithmic accuracy, can be summed to all orders and results are shown
in Fig.~\ref{Ssub} (Fig.~1 of~\cite{S4}) for the average number of
subjets.
The resummation can be seen to be extremely important for small
$y_{\mathrm{cut}}$, while the initial-state resummation is only a
relatively small correction.

\begin{figure}[t]
\centerline{\resizebox{!}{0.33\textwidth}{\includegraphics{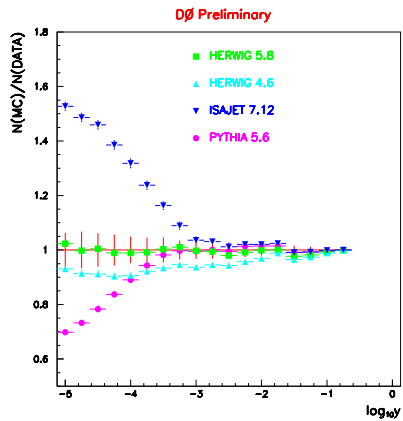}}}
\caption{The multiplicity of subjets in jets of at least 250~GeV $E_T$:
  the predictions of various Monte Carlo models are divided by the
  data.  The error bars on the \textcolor{red}{central line} are those
  of the data.  Taken from~\protect\cite{D02}.}
\label{rich}
\end{figure}

The subjet multiplicity was studied in a preliminary way by D0
in~\cite{D02} and compared with the parton shower Monte Carlo programs.
The results are shown in Fig.~\ref{rich} (Fig.~7 of~\cite{D02}).
I still find this figure amazing: on the left-hand side we are studying
250~GeV jets at a scale of less than 1~GeV where the hadronization
corrections are huge and, at least in HERWIG, the description is
perfect.  It is possible that the over-production of subjets in ISAJET
is related to the lack of colour coherence and angular ordering and that
the deficit in PYTHIA, which starts at smaller $y_{\mathrm{cut}}$ is due
to an over-estimate of the amount of `string drag' pulling soft hadrons
out of the jet, although these effects have not been studied in detail.

Subjet properties have also been studied for separate quark and gluon
jet samples using an extremely neat statistical separation\cite{D03}.
On the assumption that quark and gluon jet properties are each
independent of $\sqrt{s}$ for fixed $E_T$, the fact that the flavour mix
of jets varies strongly and that this variation is well predicted by
perturbative QCD can be used to measure their individual properties
without the need for an event-by-event tag.
\begin{figure}[t]
\centerline{\resizebox{!}{0.33\textwidth}{\includegraphics[0,0][512,512]{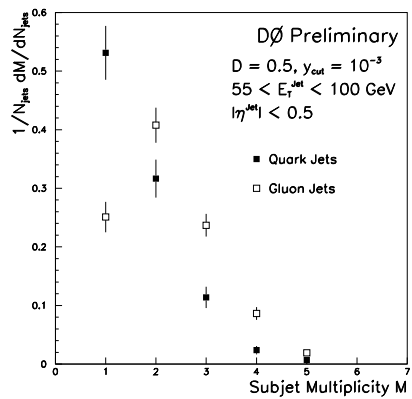}}}
\caption{Corrected subjet multiplicity in quark and gluon jets,
extracted from D0 data.  Taken from~\protect\cite{D03}.}
\label{D0sub2}
\end{figure}
The results for the subjet rates are shown in Fig.~\ref{D0sub2} (Fig.~3
of~\cite{D03}), where it can be seen that as expected gluon jets contain
a lot more activity than quark jets.  The distributions are again
well described by HERWIG.

\subsubsection{All-orders resummation for subjet rates}

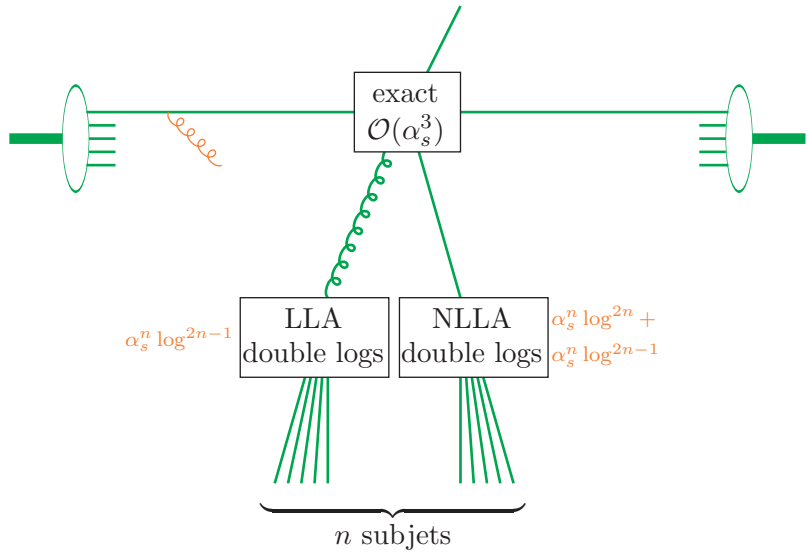
\begin{figure}[t]
\centerline{\scalebox{1}{\begin{picture}(300,200)(0,100)
\SetColor{Green}
\SetWidth{4}
\Line(0,250)(20,250)\Line(280,250)(300,250)
\SetWidth{1}
\Oval(25,250)(20,5)(0)\Oval(275,250)(20,5)(0)
\Line(30,255)(40,255)\Line(270,255)(260,255)
\Line(30,250)(40,250)\Line(270,250)(260,250)
\Line(30,245)(40,245)\Line(270,245)(260,245)
\Line(29,240)(40,240)\Line(271,240)(260,240)
\Line(29,260)(271,260)
\Line(150,260)(170,300)
\Line(150,260)(170,190)
\Gluon(150,260)(120,190){-2}{10}
\Line(170,190)(190,120)
\Line(170,190)(185,120)
\Line(170,190)(180,120)
\Line(170,190)(175,120)
\Line(170,190)(170,120)
\Line(120,190)(120,120)
\Line(120,190)(115,120)
\Line(120,190)(110,120)
\Line(120,190)(105,120)
\Line(120,190)(100,120)
\SetColor{Black}
\SetWidth{0.5}
\BBoxc(150,260)(40,30)
\Text(150,268)[c]{exact}\Text(150,254)[c]{$\mathcal{O}(\alpha_s^3)$}
\BBoxc(175,175)(56,30)
\Text(175,183)[c]{NLLA}\Text(175,169)[c]{double logs}
\BBoxc(115,175)(56,30)
\Text(115,183)[c]{LLA}\Text(115,169)[c]{double logs}
\Text(145,110)[c]{$\underbrace{\rule{100pt}{0pt}}$}
\Text(145,100)[c]{$n$ subjets}
\SetColor{Orange}
\Gluon(60,260)(80,240){-2}{4}
\Text(85,175)[r]{\Orange{$\scriptstyle\alpha_s^n\log^{2n-1}$}}
\Text(205,183)[l]{\Orange{$\scriptstyle\alpha_s^n\log^{2n}+$}}
\Text(205,169)[l]{\Orange{$\scriptstyle\alpha_s^n\log^{2n-1}$}}
\end{picture}}}
\caption{Illustration of the calculations of Refs.~\protect\cite{S4,FS}.
The primary hard parton evolves due to final state radiation and its
double logs give the leading log contribution.  It is accounted for to
next-to-leading logarithmic accuracy.  Soft initial state radiation can
also be clustered into the jet and its double logs contribute to the
next-to-leading logs.  It only needs to be calculated to leading log
accuracy since it is already one log down.  This is done by multiplying
the exact hard matrix element by a soft gluon multiplication factor.}
\label{idea}
\end{figure}

Recently the first calculation of subjet rates in hadron collisions was
performed\cite{FS}.  In general the $n$-subjet rate contains terms like
$\alpha_s^m\log^{2m}y_{\mathrm{cut}}$ at all orders of perturbation
theory $m\ge n\!-\!1$.  Together with the next-to-leading logarithms
($\alpha_s^m\log^{2m-1}y_{\mathrm{cut}}$) these can be summed to all
orders using the same trick as was used for the subjet multiplicity
in~\cite{S4}, which is illustrated in Fig.~\ref{idea}.
The evolution of the final-state jet is process-independent and can be
summed to next-to-leading log accuracy using the well-known formul\ae\
from $\mathrm{e^+e^-}$ annihilation\cite{CDFW}.  However, the
next-to-leading logs also receive a contribution from soft initial-state
radiation that happens to be close enough to the jet to be combined with
it.  As the probability of such emission depends on the full details of
the hard process through the kinematics, identities and
colour-connection of all participating partons, it seems unlikely that
this could be resummed analytically.  However, by carefully combining
the analytical result with a numerical integration of the exact matrix
element to produce one additional gluon, it is possible not only to sum
these logs to all orders, but also to automatically exactly reproduce
the ${\cal{O}}(\alpha_s)$ contribution to the one- and two-subjet rates.

\begin{figure}[t]
\centerline{\resizebox{!}{0.33\textwidth}{\includegraphics[64,237][522,593]{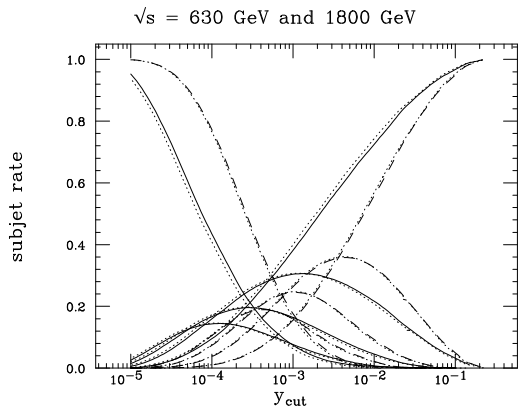}}}
\caption{$y_{\mathrm{cut}}$ dependence of the $N$ subjet rate in quark
(solid) and gluon (dashed) jets for $N=1$, 2, 3, 4~and~$\ge5$ at
$\sqrt{s}=1800$~GeV.  Also shown (dotted) are the same things at
$\sqrt{s}=630$~GeV.  Adapted from~\protect\cite{FS}.}
\label{jeff1}
\end{figure}

\begin{figure}[t]
\centerline{\resizebox{!}{0.33\textwidth}{\includegraphics[64,237][522,566]{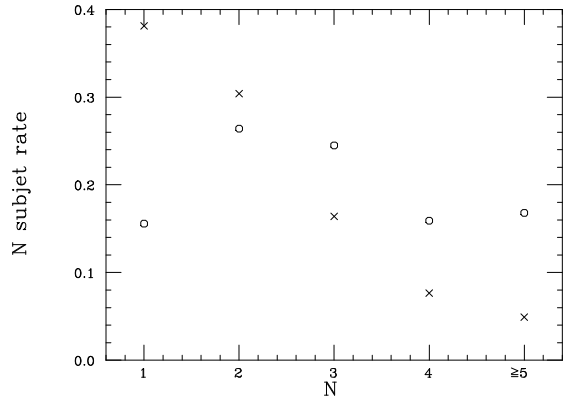}}}
\caption{Rates for $N$ subjets at $y_{\mathrm{cut}}=10^{-3}$ in quark
(crosses) and gluon (circles) jets.  Taken from~\protect\cite{FS}.}
\label{jeff2}
\end{figure}

Examples of the results are shown in Figs.~\ref{jeff1} and~\ref{jeff2}
(Figs.~9,~10 and~12 of~\cite{FS}).  The first thing to note is that
the general forms look very reminiscent of $\mathrm{e^+e^-}$
annihilation.  It is possible to separate out the contribution from
quark and gluon jets and test the hypothesis used by D0 that these are
independent of the centre-of-mass collision energy.
As can be seen in Fig.~\ref{jeff1} (Figs.~9 and~10 of~\cite{FS}) this
is the case.
The results at fixed $y_{\mathrm{cut}}$ shown in Fig.~\ref{jeff2}
(Fig.~12 of~\cite{FS}) are certainly reminiscent of D0's but, owing to
the fact mentioned earlier that they use a preclustering algorithm and
the theoretical calculation does not, direct comparison is not yet
possible.

It is worth noting that in order to extend this calculation to DIS or
photoproduction, it is necessary only to put the appropriate matrix
element into the box marked ``exact ${\cal{O}}(\alpha_s^3)$'' in
Fig.~\ref{idea}.  All the analytically-summed contributions are then
identical.

\section{Conclusion}

Precision QCD physics, and the use of jets in precision electroweak
physics, requires reliable jet definitions and reliable predictions of
the internal structure of jets.  We have reviewed a small subset of the
advances made in these areas in recent years.

\end{document}